\documentclass[twocolumn]{article}
\usepackage{authblk}
\usepackage{amsmath, amsthm, amssymb}
\usepackage{graphicx,color,psfrag}
\usepackage[normalem]{ulem}
\usepackage{hyperref}
\usepackage{cleveref}
\usepackage{caption}
\usepackage{subcaption}
\usepackage{thmtools,mathtools}
\declaretheoremstyle[notefont=\bfseries,notebraces={}{},%
    headpunct={},postheadspace=1em]{mystyle}

\DeclarePairedDelimiter\abs{\lvert}{\rvert}%
\DeclarePairedDelimiter\norm{\lVert}{\rVert}%
\makeatletter
\let\oldabs\abs
\def\abs{\@ifstar{\oldabs}{\oldabs*}}
\let\oldnorm\norm
\def\norm{\@ifstar{\oldnorm}{\oldnorm*}}
\makeatother
\usepackage{color}
\newcommand{\Lfun}[2]{{\cal L}^{(#1)}_{#2}}

\DeclareMathAlphabet{\gcal}{OMS}{cmsy}{m}{n}

\graphicspath{{./}}
\begin{document}
\title{Scanning Tunneling Thermometry}
\author[1,2]{Abhay Shastry\thanks{Corresponding Author: abhayshastry@email.arizona.edu}}
\author[1,3]{Sosuke Inui}
\author[1]{Charles\ A.\ Stafford}
\affil[1]{Department of Physics, University of Arizona, Tucson, AZ 85721, USA}
\affil[2]{Chemical Physics Theory Group, Department of Chemistry, University of Toronto, 80 St. George Street, Toronto, ON M5S 3H4, Canada}
\affil[3]{Department of Physics, Osaka City University, Sugimoto 3-3-138, Sumiyoshi-Ku, Osaka 558-8585, Japan}
\maketitle
\begin{abstract}
{\bf Temperature imaging of nanoscale systems is a fundamental problem which has myriad potential technological applications. 
For example, nanoscopic cold spots can be
used for spot cooling electronic components while hot spots
could be used for precise activation of chemical or biological reactions. More fundamentally, imaging the temperature fields
in quantum coherent conductors can provide a wealth of information on heat flow and dissipation at the smallest scales.
However, despite significant technological advances, the spatial resolution of temperature imaging remains in the
few nanometers range.
% Contribution from different degrees of freedom make it
% very difficult to image the {\em electronic} temperature in a nanoscale conductor.
Here we propose a method to
map electronic temperature variations in operating nanoscale conductors by
relying solely upon electrical tunneling current measurements. The {\em scanning tunneling thermometer}, 
owing to its operation in the tunneling regime,  
would be capable of mapping sub-angstrom temperature variations, thereby enhancing
the resolution of scanning thermometry by some two orders of magnitude.}
\end{abstract}

Thermal imaging of nanoscale systems is of crucial importance not only due to its potential
to enable future technologies, but
also because it can greatly enhance our understanding of heat transport at the smallest scales. In recent years, nanoscale thermometry
has been used in a wide range of fields \cite{Brites2012} including thermometry in a living cell \cite{Kucsko2013}, local control of chemical
reactions \cite{Jin2011} and temperature mapping of operating electronic devices \cite{Mecklenburg2015}. Various studies utilize radiation-based
techniques such as Raman spectroscopy \cite{Reparaz2014}, fluorescence in nanodiamonds \cite{Kucsko2013,Neumann2013}
and near-field optical microscopy \cite{Teyssieux2007}. 
The spatial resolution of these radiation-based techniques is limited due to optical
diffraction and, to overcome this drawback, scanning probe techniques have seen a flurry of activity in recent years
\cite{Gomes2015}. However, despite their remarkable progress, the spatial resolution remains in the $10\ \rm{nm}$ range.
A key obstacle to achieving high spatial resolution in scanning probe thermometry has 
been the fundamental difficulty in designing a thermal probe that exchanges heat with the system of interest
but is thermally isolated from the environment.

Since temperature and voltage are both fundamental thermodynamic observables, it is instructive to draw the sharp contrast
that exists between
the measurement of these two quantities at the nanoscale.
Scanning tunneling potentiometry (STP) \cite{Muralt1986} is a mature technology and can map local voltage variations
with sub-angstrom spatial resolution by operating in the tunneling regime.
STP has been used to map the local voltage variations
in the vicinity of individual scatterers, interfaces or boundaries \cite{Briner1996,Ramaswamy1999,Wang2013,Clark2014,Willke2015},
providing direct observations of the Landauer dipole \cite{Landauer1957,Buttiker89}. 
STP has been a useful tool in disentangling
different scattering mechanisms \cite{Willke2015} and can map local potential variations 
due to quantum interference effects \cite{Wang2013,Clark2014}. Similarly, local temperature variations 
due to quantum interference effects have been theoretically predicted for various nanosystems out of equilibrium \cite{Dubi2009b,Bergfield2013demon,Meair14,Bergfield2015}
but have hitherto remained outside the reach of experiment. 

Scanning thermal microscopy \cite{Williams1986} (SThM) relies on the measurement of a heat-flux signal that can be sensed, 
e.g., by a calibrated thermocouple or an electrical resistor \cite{Majumdar1999}. A good thermal contact between
the tip and sample is needed for an appreciable heat flux and generally implies a measurement in the contact regime,
thus limiting the spatial resolution. 
Despite the recent progress in adressing contact-related issues in SThM \cite{Gomes2015,Menges2016b}, 
the best spatial resolution is presently $\sim7\ \rm{nm}$ \cite{Menges2016}. 

It is well known that, outside equilibrium, the temperatures
of different degrees of freemdom (e.g.\ phonon, photon, electron) do not coincide \cite{Casas-Vazques2003};
Existing SThM schemes cannot distinguish between the 
contributions of the different degrees of freedom to the heat-flux.
A number of nanoscopic devices operate in the elastic transport regime where the electron and phonon degrees of freedom
are completely decoupled and, consequently, the distinction between their temperatures becomes extremely important \cite{Stafford2016}.

From a fundamental point of view, a thermometer is a device that equilibrates locally with the system of interest and has some 
temperature-dependent physical property (e.g.\ resistance, thermopower, mass density) which can be measured;
the temperature measurement seeks to find the condition(s) under which the thermometer is in local thermodynamic equilibrium with
the system of interest and concurrently infers the thermometer's temperature by relying upon those temperature-dependent physical
properties. 
Ideally, the measurement apparatus must not substantially disturb the state of the system of interest \cite{Shastry2016a}.
SThM schemes, by relying on heat fluxes in the contact regime, may alter the state of a small system.
We propose here a noninvasive
thermometer whose local equilibration can be inferred by the measurement of electrical tunneling currents alone.
In particular, we find that the conditions required for the local equilibration of the {\em scanning tunneling thermometer} (STT)
are completely determined by (a) the conductance and thermopower
which are both measured using the tunneling current and (b) the bias conditions of the conductor defined 
by the voltages and temperatures of the contacts.

Our proposed method relies solely upon electrical
measurements made in the tunneling regime and provides a measurement of the {\em electronic} temperature decoupled
from all other degrees of freedom. We predict a dramatic enhancement of the spatial resolution by more than two orders
of magnitude, thereby bringing thermometry to the sub-angstrom regime.
The method is valid for systems obeying the Wiedemann-Franz (WF) law \cite{AshcroftAndMerminBook} which relates
the electrical ($G$) and thermal conducatances ($\kappa$) in a material-independent way $\kappa/G = \pi^{2}k_{B}^{2}T/3e^{2}$.
The WF law was first observed in bulk metals over 150 years ago and has been verified in a large number of nanoscale conductors.
Most recently, it has been validated in atomic contact junctions \cite{Mosso2017, Cui2017} which
represent the ultimate limit of miniturization of electronic conductors.

\section*{Temperature measurement}

We note a crucial, but often overlooked, theoretical point pertaining to the imaging of temperature fields on a nonequilibrium conductor.
The prevailing paradigm for temperature and voltage measurements is the following \cite{Engquist81}: 
(i) a voltage is measured by a probe (voltmeter) when in electrical equilibrium with the sample
and (ii) a temperature is measured by a probe (thermometer) 
when in thermal equilibrium with the sample. We refer to this definition as the Engquist-Anderson (EA)
definition.

The fact that the EA definition implicitly ignores
thermoelectric effects was pointed out by Bergfield and Stafford \cite{Bergfield2013demon,Bergfield2014}, 
and a notion of a joint probe
was put forth by requiring {\em both} electrical and thermal equilibrium with the sample.
It is quite easy to understand this intuitively:
A temperature probe lacking local electrical equilibration with the sample develops a temperature bias at the probe-sample junction due to the Peltier
effect; similarly, a voltage probe lacking local thermal equilibration with 
the sample develops a voltage bias at the probe-sample junction due to
the Seebeck effect. These errors \cite{Bergfield2014} can be quite large for systems with large thermoelectric responses.
A temperature probe therefore has to remain in thermal {\em and} electrical equilibrium with the nonequilibrium sample
\cite{Bergfield2013demon,Meair14,Stafford2016,Shastry2016a,Bergfield2014,Shastry2015}, thereby ensuring true
{\em thermodynamic} equilibrium of the measurement apparatus.

The joint probe measurement was made mathematically rigorous in a recent study by
Shastry and Stafford \cite{Shastry2016a} where it was shown that the 
solution to the probe equilibration problem always exists and is unique, arbitrarily
far from equilibrium and with arbitrary interactions within the quantum system. Moreover, it was shown that the EA definition 
is provably nonunique: The value measured by the EA thermometer depends quite strongly
on its voltage and, conversely, the value measured by the EA voltmeter depends on its temperature. These results are 
intimately connected to the second law of thermodynamics and expose the fatal flaw in the EA definition: the measurement
apparatus (thermometer or voltmeter) has to remain in {\em thermodynamic} equilibrium,
i.e., {\em electrical} and {\em thermal} equilibrium, with the system of interest which it probes locally. 
Simply stated, an open system of electrons (a conductor) exchanges both charge {\em and} heat.
% A measurement of temperature therefore has to involve both electrical and thermal equilibration 
% simply due to the fact that electrons carry both charge and heat. 
We therefore write
\begin{equation}
I_{p}=0=J_{p},
\label{Measurement}
\end{equation}
for the simultaneous vanishing of the electric current $I_{p}$ and the {\em electronic} contribution to the heat current $J_{p}$
flowing into the probe $p$.
The above equation determines the conditions under which a local thermodynamic equilibrium is established between
the probe (STT) and the nonequilibrium system of interest.

\section*{Temperature from tunneling currents}

The probe currents depend linearly on the temperature and voltage gradients for transport
within the linear response regime:
\begin{equation}
\large
\begin{pmatrix}
I_{p}\\
J_{p}
\end{pmatrix}=\sum_{\alpha}\begin{pmatrix}
\Lfun{0}{p\alpha} & \Lfun{1}{p\alpha}\\
\Lfun{1}{p\alpha} & \Lfun{2}{p\alpha}
\end{pmatrix}\begin{pmatrix}
V_{\alpha}-V_{p}\\
\frac{T_{\alpha}-T_{p}}{T_{0}}
\end{pmatrix},
\normalsize
\label{LinearResponse}
\end{equation}
where the $\Lfun{\nu}{p\alpha}$ are the Onsager linear response coefficients evaluated at the equilibrium temperature $T_{0}$
and chemical potential $\mu_{0}$. 
$\Lfun{0}{p\alpha}$ is the electrical conductance ($\Lfun{0}{p\alpha} = G_{p\alpha}$) between the probe $p$ and contact $\alpha$.
$\Lfun{1}{p\alpha}$ is related to the thermopower ($S_{p\alpha}$) and electrical conductance 
(${\Lfun{1}{p\alpha}}= -T_{0} S_{p\alpha}{G_{p\alpha}}$). Finally, $\Lfun{2}{p\alpha}$ is related to the 
thermal conductance ($\Lfun{2}{p\alpha}= T_{0} \kappa_{p\alpha}$) 
up to leading order in the Sommerfeld series \cite{AshcroftAndMerminBook} (see also the Supplementary Information). 

We solve for the temperature of the STT in Eq.\ (\ref{Measurement}) and find \cite{Meair14}
\begin{equation}
\begin{aligned}
&\frac{T_{p}^{(\rm{Exact})}}{T_{0}}
= \frac{\sum_{\beta}\Lfun{0}{p\beta}\sum_{\alpha}\Lfun{1}{p\alpha}V_{\alpha}
-\sum_{\beta}\Lfun{1}{p\beta}\sum_{\alpha}\Lfun{0}{p\alpha}V_{\alpha}}
{\sum_{\beta}\Lfun{2}{p\beta}\sum_{\alpha}\Lfun{0}{p\alpha}-(\sum_{\alpha}\Lfun{1}{p\alpha})^2}\\
&+ \frac{1}{T_{0}}\frac{\sum_{\beta}\Lfun{0}{p\beta}\sum_{\alpha}\Lfun{2}{p\alpha}{T_{\alpha}}- \sum_{\beta}\Lfun{1}{p\beta}
\sum_{\alpha}\Lfun{1}{p\alpha}{T_{\alpha}}}
{\sum_{\beta}\Lfun{2}{p\beta}\sum_{\alpha}\Lfun{0}{p\alpha}-(\sum_{\alpha}\Lfun{1}{p\alpha})^2}.
\label{LinearResponseTemperature}
\end{aligned}
\end{equation}
Here $T_{p}^{(\rm{Exact})}$ denotes the exact solution to the equilibration of the STT, i.e., Eq.\ (\ref{Measurement}),
within the linear response regime where the currents are expressed by Eq.\ (\ref{LinearResponse}).

\begin{figure*}[tbh]
\centering
\captionsetup{justification=raggedright,
singlelinecheck=false
}

\includegraphics[width=5in]{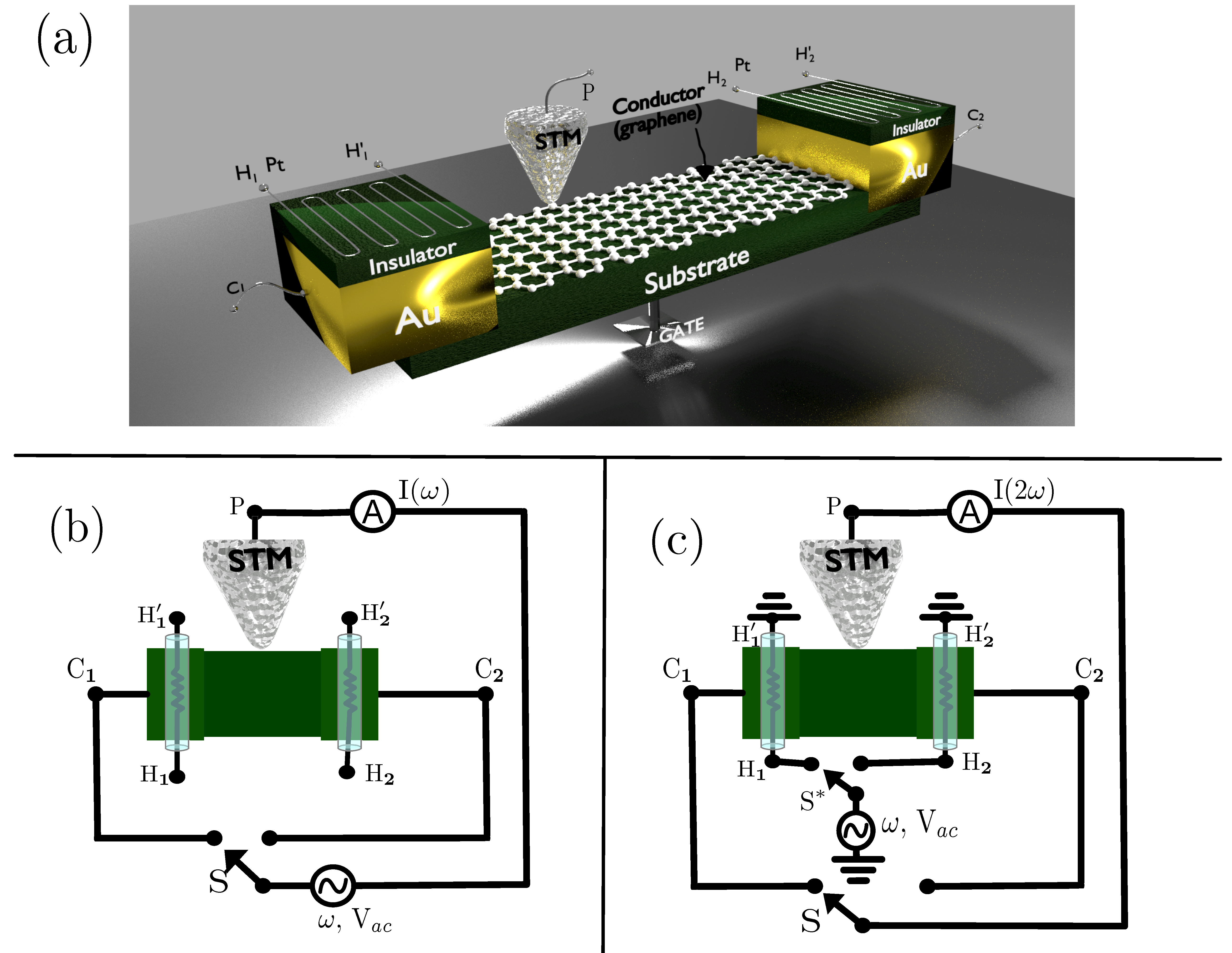}
\caption{(a) Schematic depiction of the system and measurement apparatus. An STM tip scans the surface of a nanoscale
conductor at a fixed height. The conductor sits on top of a substrate which may be gated. Two gold contacts $C_{1}$ and $C_{2}$
are connected to the conductor on either side. A Pt heater ($H_{1}H'_{1}$ and $H_{2}H'_{2}$) 
sitting atop each gold contact, and electrically insulated from it, allows
one to modulate the temperature of the gold contacts. (b) The conductance circuit which measures the coefficient $\Lfun{0}{p\alpha}$
for each contact $\alpha$ selected using switch S. (c) The thermoelectric circuit which measures
the coefficient $\Lfun{1}{p\alpha}$ for each contact $\alpha$. Switch S* activates the heater in the corresponding contact $\alpha$
selected by switch S.
}
\label{fig:Setup}
\end{figure*}

% They are related to the electrical conductance ($G_{p\alpha}= \Lfun{0}{p\alpha}$), thermopower 
% ($S_{p\alpha}= -{\Lfun{1}{p\alpha}}/{ T_{0}\Lfun{0}{p\alpha}}$) and thermal conductance 
% [$\kappa_{p\alpha}= \big(\Lfun{2}{p\alpha}  -{(\Lfun{1}{p\alpha})^{2}}/{\Lfun{0}{p\alpha}}\big)/{T_{0}}$].
% \begin{equation}
% \begin{aligned}
% G_{p\alpha}(\mu_{0},T_{0})=&\ \Lfun{0}{p\alpha}(\mu_{0},T_{0})\\
% S_{p\alpha}(\mu_{0},T_{0})=& -\frac{\Lfun{1}{p\alpha}(\mu_{0},T_{0})}{ T_{0}\Lfun{0}{p\alpha}(\mu_{0},T_{0})}\\
% \kappa_{p\alpha}(\mu_{0},T_{0})=& \frac{1}{T_{0}}\bigg(\Lfun{2}{p\alpha}(\mu_{0},T_{0}) 
% -\frac{(\Lfun{1}{p\alpha})^{2}}{\Lfun{0}{p\alpha}}\bigg).
% \end{aligned}
% \label{definitions:GSK}
% \end{equation}
Eq.\ (\ref{LinearResponse}) suggests that $\Lfun{0}{p\alpha}$ and $\Lfun{1}{p\alpha}$ can be measured
using the tunneling current $I_{p}$, whereas $\Lfun{2}{p\alpha}$ appears only in the expression for the heat current $J_{p}$
and would generally involve the measurement of a heat-flux-related signal. However, for systems obeying the WF law,
we may simply relate $\Lfun{0}{p\alpha}$ and $\Lfun{2}{p\alpha}$ using
\begin{equation}
\Lfun{2}{p\alpha}= \frac{\pi^{2}k_{B}^{2}T_{0}^{2}}{3e^{2}}\Lfun{0}{p\alpha},
\label{WF}
\end{equation}
valid up to leading order in the Sommerfeld series. This allows us to infer $\Lfun{2}{p\alpha}$ using the WF law given by 
Eq.\ (\ref{WF}) and we obtain
\begin{equation}
\begin{aligned}
\frac{T_{p}^{\rm(WF)}}{T_{0}}=& \frac{3e^{2}}{\pi^{2}k_{B}^{2}T_{0}^{2}}\bigg(\frac{\sum_{\alpha}\Lfun{1}{p\alpha}V_{\alpha}}{\sum_{\alpha}\Lfun{0}{p\alpha}}
-\frac{\sum_{\alpha}\Lfun{1}{p\alpha}}{\sum_{\alpha}\Lfun{0}{p\alpha}}
\frac{\sum_{\beta}\Lfun{0}{p\beta}V_{\beta}}{\sum_{\beta}\Lfun{0}{p\beta}}\bigg)
\\&+
\frac{\sum_{\alpha}\Lfun{0}{p\alpha}{T_{\alpha}}}{{T_{0}}\sum_{\alpha}\Lfun{0}{p\alpha}},
\label{OurApproximation}
\end{aligned}
\end{equation}
valid up to leading order in the Sommerfeld series.
% \begin{equation}
% \Lfun{2}{p\alpha}\overset{\rm{WF\ law}}{\rightarrow} 
% \frac{\pi^{2}}{3}k_{B}^{2}T_{0}^{2}\Lfun{0}{p\alpha} \implies T_{p}^{(\rm{Exact})}\rightarrow T_{p}^{(\rm{WF})}.
% \label{WFApproximatedSolution}
% \end{equation}

$T_{p}^{(\rm{WF})}$ requires only the measurement of $\Lfun{0}{p\alpha}$ and $\Lfun{1}{p\alpha}$, or equivalently,
the electrical conductance and thermopower, and lends itself to a simple interpretation: 
The first term in Eq.\ (\ref{OurApproximation})
is the thermoelectric contribution whereas the second term is the thermal contribution. 
The second-order corrections in the Sommerfeld series are typically very small
\begin{equation}
T_{p}^{(\rm{WF})}= T_{p}^{(\rm{Exact})}\bigg(1 +  \mathcal{O}\big((k_{B}T_{0}/\Delta)^{2} \big)\bigg),
\label{Corrections}
\end{equation}
where the characteristic energy
scale of the problem $\Delta$ is typically much larger than the thermal energy set by $k_{B}T_{0}$:
e.g., $\Delta=\epsilon_{F}$, the Fermi energy, for bulk systems and for a tunneling probe $\Delta$
is of the order of the work function. The breakdown of the Wiedemann-Franz law has been reported in various nanoscale
systems. The characteristic energy scale $\Delta$ in such cases is comparable
to the thermal energy thereby leading to large errors in the Sommerfeld series expansion in Eq.\ (\ref{Corrections}). 
In graphene, the breakdown of the WF law was reported in Ref.\ \cite{Crossno2016}. Here, the local chemical potential was
tuned (via local doping) such that it is smaller than the thermal energy thereby creating the so-called Dirac fluid. 
Such systems show a decoupling of charge and heat currents, making it impossible to measure heat currents 
through electrical means. Although our results apply to a broad array of nanoscale conductors, 
they do not apply to systems prepared in this manner.

It is clear from Eq.\ (\ref{OurApproximation}) that the measurement of (a) conductance $\Lfun{0}{p\alpha}$ and 
the thermoelectric coefficient $\Lfun{1}{p\alpha}$ along with the 
(b) known bias conditions of the system $\{ V_{\alpha},T_{\alpha}\}$
completely determine the conditions under which the STT is in local thermodynamic equilibrium
with the system.

\section*{Experimental Implementation}

The temperature measurement involves two circuits: (I) The conductance circuit which measures the
electrical conductance $\Lfun{0}{p\alpha}$ and (II) The thermoelectric circuit which measures the
thermoelectric response coefficient $\Lfun{1}{p\alpha}$, as shown in Fig.\ \ref{fig:Setup} (b) and (c) respectively.
The STT involves operating the tip of a scanning tunneling microscope (STM) at a constant height above the surface
of the conductor in the tunneling regime. The circuit operations (I) and (II) are described below. 

(I) {\em The conductance circuit} involves a closed circuit of the probe and the contact $\alpha$. 
All contacts and the probe are held at  
the equilibrium temperature $T_{\alpha}=T_{p}=T_{0}$. An ac voltage $V(\omega)$
is applied at the probe-contact junction $V(\omega)= V_{p}-V_{\alpha}$
and the resulting tunneling current $I_{p}(\omega)$ is recorded
using standard lock-in techniques. The STM tip
is scanned along the surface. A switch disconnects all contacts except $\alpha$ and the tunneling
current is therefore
\begin{equation}
\begin{aligned}
I_{p}&= \Lfun{0}{p\alpha}(V_{p}-V_{\alpha})= -I_{\alpha},\\
I_{p}(\omega)&=\Lfun{0}{p\alpha}V(\omega).
\label{eq:ConductanceCircuit}
\end{aligned}
\end{equation}
The procedure is repeated for all the contacts $\alpha$ by toggling the switch S
shown in Fig.\ (\ref{fig:Setup}b) and a scan is obtained for each probe-contact junction.
This completes the measurement of the conductance $\Lfun{0}{p\alpha}$ for all the contacts $\alpha$.

(II) {\em The thermoelectric circuit} involves a (i) closed circuit of the probe and contact $\alpha$, 
which is the same as the conductance circuit without the voltage source, and
(ii) an additional circuit which induces time-modulated temperature variations in contact $\alpha$; 
% The temperature modulations in contact $\alpha$ are induced by a Pt resistance thermometer fabricated on top 
% of an insulator above contact $\alpha$ (for electrical insulation)
% as shown in fig.\ (\ref{fig:Circuit}b); 
An ac current at frequency $\omega$
induces Joule heating in the Pt resistor at frequency $2\omega$
and results in a temperature modulation $T_{\alpha}= T_{0} + \Delta T_{\alpha}(2\omega)$
in the contact $\alpha$. The probe is held at the equilibrium temperature $T_{p}= T_{0}$.
The resulting tunneling current $I_{p}(2\omega)$, at frequency $2\omega$, is recorded using standard lock-in techniques.
The STM tip is scanned along the surface at the same points as before. A switch disconnects all contacts except $\alpha$
and the tunneling current is
\begin{equation}
\begin{aligned}
I_{p}&= \Lfun{1}{p\alpha}\frac{(T_{\alpha}-T_{p})}{T_{0}} = -I_{\alpha}\\
I_{p}(2\omega)&=\Lfun{1}{p\alpha}\frac{\Delta T_{\alpha}(2\omega)}{T_{0}}.
\end{aligned}
\label{eq:ThermoelectricCircuit}
\end{equation}
The procedure is repeated for all the contacts $\alpha$ by toggling the switches S and S*
shown in Fig.\ (\ref{fig:Setup}c) and a scan is obtained for each probe-contact junction.
Note that the switch S* must heat the Pt resistor in the same contact $\alpha$ for
which the probe-contact tunneling current is measured.
This completes the measurement of the thermoelectric coefficient $\Lfun{1}{p\alpha}$ for all the contacts $\alpha$.
\begin{figure*}[tbh]
\centering
\captionsetup{justification=raggedright,
singlelinecheck=false
}
{\includegraphics[width=3.5in]{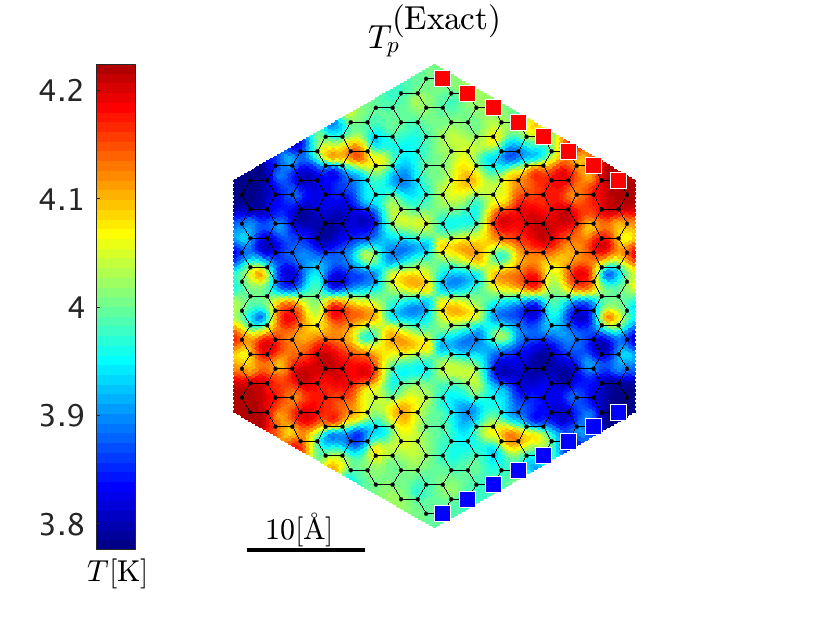}}\hspace{-0.7in}
{\includegraphics[width=3.5in]{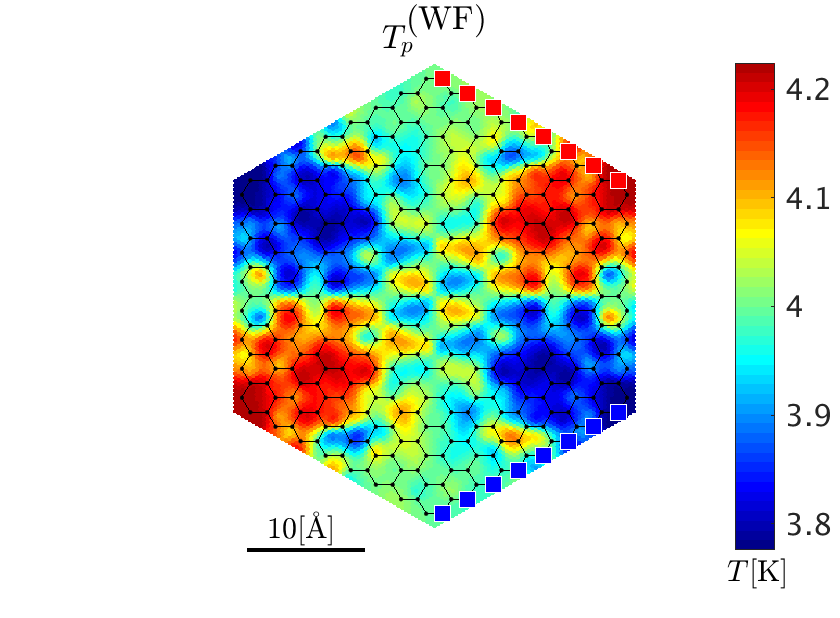}}
\caption{Temperature variations on a hexagonal graphene flake with an application of a symmetrical temperature bias 
$T_{\rm{red}}-T_{\rm{blue}}=0.5\,\rm{K}$,
where the red (hot) and blue (cold) squares indicate the sites coupled to the contacts; $T_{0}=4\,\rm{K}$. The left panel shows the exact linear response
solution given in Eq.\ \ref{LinearResponseTemperature} while the right panel shows the WF solution given
by Eq.\ (\ref{OurApproximation}). The same temperature scale is used for both panels.
}
\label{fig:TemperatureBias}
\end{figure*}

Heating elements have been fabricated in the contacts previously \cite{Tsutsui2012}. Any system where one may induce Joule heating
can be used as the heating element (instead of Pt) in the circuit. 
For example, another flake of graphene could be used as a heating element as long
as it is calibrated accurately.
The voltage modulation frequency in the heating elements $\omega\ll1/\tau$, where $\tau$ is the thermal time constant of the contacts,
so that the contact may thermalize with the heating element. Typically, $\tau$ is of the order of tens of nanoseconds (cf. methods
in \cite{Bergfield2013demon}).
We discuss the calibration of the contact temperature $T_{\alpha}= T_{0} + \Delta T_{\alpha}$ in the Supplementary Information.
The thermoelectric response of the nanosystem may be quite sensitive to the gate voltage which is also discussed in the Supplementary
Information.

\begin{figure*}[tbh]
\centering
\captionsetup{justification=raggedright,
singlelinecheck=false
}
{\includegraphics[width=3.5in]{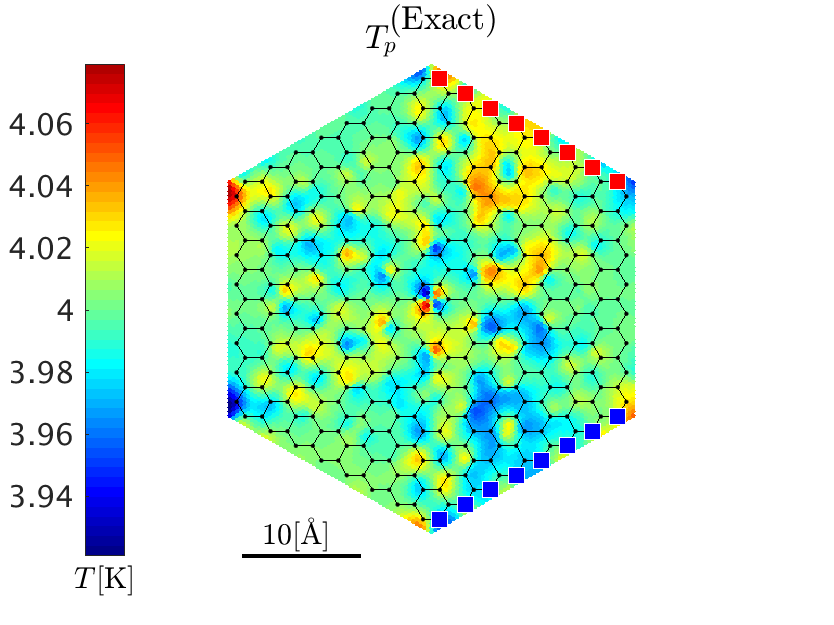}}\hspace{-0.8in}
{\includegraphics[width=3.5in]{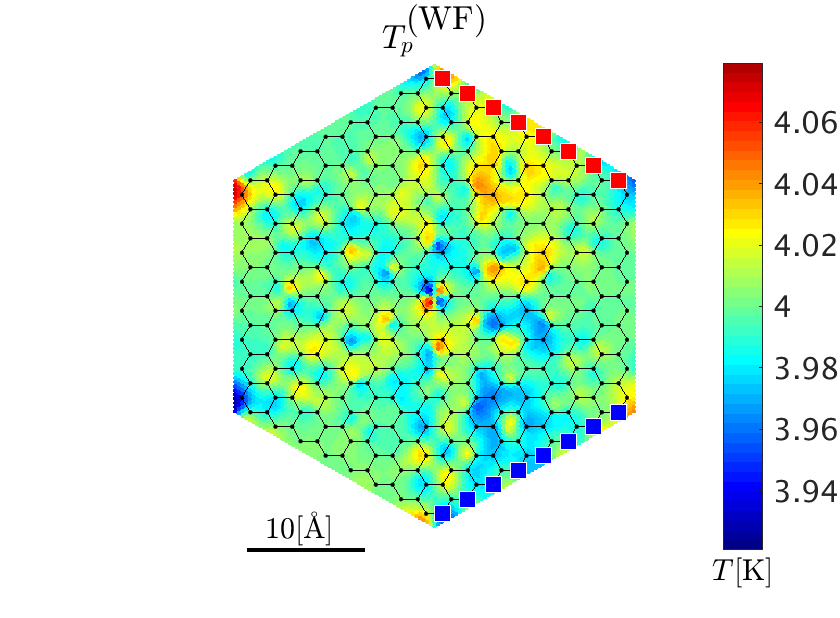}}
\caption{Temperature variations on a hexagonal graphene flake with an application of a voltage bias 
$V_{\rm{blue}}-V_{\rm{red}}=k_{B}T_{0}/e= 0.34\ \rm{mV}$,
where the red and blue squares indicate the sites coupled to the contacts; $T_{0}=4\,\rm{K}$. The left panel shows the exact linear response
solution given in Eq.\ \ref{LinearResponseTemperature} while the right panel shows the approximate solution obtained by employing the
WF law given in Eq.\ \ref{OurApproximation}. The same temperature scale is used for both panels.
}
\label{fig:VoltageBias}
\end{figure*}

\section*{Results}

We present model temperature measurements for a hexagonal graphene flake under 
(a) a thermal bias and (b) a voltage bias. 
The measured temperature, for a combination of thermal and voltage biases, would simply be a linear
combination of the two scenarios (a) and (b) in the linear response regime (under identical
gating conditions). Therefore, we present the two
cases separately but we note that the gate voltages are not the same for the two scenarios that we present here. 
The voltage bias case has been gated differently so as to enhance the thermoelectric response of the system.
% Our choice of graphene is based on the fact that its transport properties can be tuned by an appropriate
% choice of the gate voltage
We show the temperature measurement
for (a) the thermal bias case in Fig.\ \ref{fig:TemperatureBias} and (b) the voltage bias case in Fig.\ \ref{fig:VoltageBias}.
The two panels in Figs.\ \ref{fig:TemperatureBias} and \ref{fig:VoltageBias} compare 
(1) the temperature measurement $T_{p}^{(\rm{Exact})}$
obtained from the exact solution [given by Eq.\ (\ref{LinearResponseTemperature})] and 
(2) the temperature measurement $T_{p}^{(\rm{WF})}$ obtained from our method [given by Eq.\ (\ref{OurApproximation})]
which relies on the WF law.

Graphene is highly relevant for future electronic technologies and provides a versatile system
whose transport properties can be tuned by an appropriate choice of the gate voltage --- we therefore
illustrate our results for graphene. The method itself is valid for any system obeying the WF law.
% The temperature variations arising from a voltage bias are due to the thermoelectric term
% in Eq.\ (\ref{OurApproximation}) which may be tuned by gating the system \cite{Bergfield2014}. 
The thermoelectric response
coefficient $\Lfun{1}{p\alpha}\sim T_{0}^{2}$ has a quadratic suppression at low temperatures and its measurement
from Eq.\ (\ref{eq:ThermoelectricCircuit}) depends crucially on the choice of gating especially at cryogenic 
operating temperatures $\sim 4\ \rm{K}$ since the resulting tunneling current must be experimentally resolvable.
In graphene, we find that the electrical tunneling currents arising from its thermoelectric response 
are resolvable even at cryogenic
temperatures when the system is gated appropriately and, owing to the fact that a number of STM experiments 
are conducted at low temperatures, we present our results for $T_{0}= 4\ \rm{K}$. 
Higher operating temperatures result in a higher tunneling current in Eq.\ (\ref{eq:ThermoelectricCircuit})
and gating would therefore be less important. 

The $\pi$-electron system of graphene is described using the
tight-binding model whose basis states are $2p_{z}$ orbitals
at each atomic site of carbon. The STT is modeled as an atomically
sharp Pt tip operating at a constant height of $3\ \rm{\AA}$ above the plane of the carbon nuclei. The details of the
graphene Hamiltonian as well as the probe-system tunnel coupling are presented in the Methods section.
The atomic sites of graphene which are
coupled to the contacts are indicated in Figs.\ \ref{fig:TemperatureBias} and \ref{fig:VoltageBias} by either 
a red or blue square. The chemical potential and temperature of the two contacts (red and blue) set the
bias conditions for the problem. The coupling to the two contacts is symmetrical and the coupling strength
for all the coupling sites (red or blue) is taken as $\Gamma= 0.5\ \rm{eV}$.
Additional details regarding the gating and the tunneling currents 
are included in the Supplementary Information.

% Wrong Sentence: The thermoelectric coefficient $\Lfun{1}{p\alpha} \sim T_{0}^{2}$ has a quadratic supression at low temperatures
%and the gating is crucial to produce appreciable temperature variations across the sample.
Fig.\ \ref{fig:TemperatureBias} shows the variations in temperature for a symmetrical ($T_{\rm{red}}+T_{\rm{blue}}=2 T_{0}$)
temperature bias $T_{\rm{red}}-T_{\rm{blue}}=0.5\ \rm{K}$. 
The agreement between $T^{\rm{(Exact)}}_{p}$ and $T^{\rm{(WF)}}_{p}$ given by Eqs.\ (\ref{LinearResponseTemperature}) 
and (\ref{OurApproximation}) respectively is excellent. The gating has been chosen to be $\mu_{0}=-2.58\ \rm{eV}$ with respect
to the Dirac point in graphene.
The same temperature scale is used for both the panels 
in Fig.\ \ref{fig:TemperatureBias}.
% Discrepencies are due to higher-order contributions in the Sommerfeld series [cf. Eq.\ (\ref{Corrections})] and
% are extremely small. 
The temperature variations in $T^{\rm{(WF)}}_{p}$ are solely the result
of the temperature bias and are given by the second term in Eq.\ (\ref{OurApproximation}). Therefore, we require
only the measurement of the conductances $\Lfun{0}{p\alpha}$ for the temperature measurement under these bias conditions.
We consider a contact-tip voltage modulation of $1\ \rm{mV}$ for the measurement of the conductance. The resulting
tunneling currents are of the order of $10\ \rm{nA}$ with a maximum tunneling current of about $30\ \rm{nA}$. We present
the details in the Supplemental Information.

Fig.\ \ref{fig:VoltageBias} shows the variations in temperature for a 
voltage bias of $V_{\rm{blue}}-V_{\rm{red}}=k_{B}T_{0}/e$, with $T_{0}=4\ \rm{K}$,
so that the transport is within the linear response regime. The gating for this case has been chosen to be 
$\mu_{0}=-2.28\ \rm{eV}$ such that there is an enhanced thermoelectric effect. The tunneling currents
from the thermoelectric circuit, under these gating conditions, are of the order of $100\ \rm{pA}$ with 
a maximum tunneling current of about $I = 150\ \rm{pA}$ and are resolvable under standard lock-in techniques.
The variation of the contact temperature is
taken to be $\Delta T = (10\%)\ T_{0}$ with $T_{0}= 4\ \rm{K}$.
The resolution of the tunneling current is an important point especially for the measurement of the thermoelectric
response coefficient $\Lfun{1}{p\alpha}$ and has been covered in greater detail in the Supplementary Information.
The same temperature scale is used for both panels in Fig.\ \ref{fig:VoltageBias} 
and there is excellent agreement between $T^{\rm{(Exact)}}_{p}$ and $T^{\rm{(WF)}}_{p}$. 
The temperature variations shown here are solely the result
of the voltage bias and are given by the first term in Eq.\ (\ref{OurApproximation}). Since the variations are purely due to
the thermoelectric effect, the EA definition would have noted no temperature variations at all.

% In both Figs.\ \ref{fig:TemperatureBias} and \ref{fig:VoltageBias}, the hottest spots recorded in $T^{\rm{(WF)}}_{p}$
% are slightly hotter than the ones given by the exact solution $T^{\rm{(Exact)}}_{p}$; likewise,
% the coldest spots in our method is slightly colder than the ones given by the exact solution. 
The disagreement between the exact solution and our method are due to higher-order contributions in the
Sommerfeld series which are extremely small [cf.\ Eq.\ (\ref{Corrections})]. An explicit expression for the first Sommerfeld correction
in the WF law has been derived in the Supplementary Information.
The discrepency
between $T^{\rm{(WF)}}_{p}$ and $T^{\rm{(Exact)}}_{p}$ defined by $|{T^{\rm{(WF)}}_{p}-T^{\rm{(Exact)}}_{p}}|/T^{\rm{(Exact)}}_{p}$
is less than $0.01\%$ for the temperature bias case in Fig.\ \ref{fig:TemperatureBias}, 
whereas their discrepency for the voltage bias case in Fig.\ \ref{fig:VoltageBias} is
less than $0.2\%$.

\section*{Conclusion}

It has proven extraordinarily challenging to achieve high spatial resolution in thermal measurements. 
A key obstacle has been the fundamental difficulty in designing a thermal probe that exchanges heat with the system of interest
but is thermally isolated from the environment.
We propose circumventing this seemingly intractable problem by inferring thermal signals using purely electrical measurements.
The basis of our approach is the Wiedemann-Franz law relating the thermal and electrical currents flowing between a probe and the 
system of interest.

We illustrate this new approach to nanoscale thermometry with simulations of a scanning tunneling probe of a model nanostructure
consisting of a graphene flake under thermoelectric bias.  We show that the local temperature inferred from a sequence of purely 
electrical measurements agrees exceptionally well with that of a hypothetical thermometer coupled 
locally to the system and isolated from the environment. Moreover, our measurement provides the {\em electronic} temperature
decoupled from all other degrees of freedom and can therefore be a vital tool to characterize nonequilibrium device performance.
Our proposed {\em scanning tunneling thermometer} exceeds the spatial
resolution of current state-of-the-art thermometry by some two orders of magnitude. 

\section*{Methods}

\subsection*{System Hamiltonian}

The $\pi$-electron system of graphene is described within
the tight-binding model,
$H_{\rm gra}= \sum\limits_{<i,j>}t_{ij}d_{i}^{\dagger}d_{j} +{\rm h.c}$, 
with nearest-neighbor hopping matrix element $t_{ij}=-2.7$eV.
The coupling of the system with the contact reservoirs is described by the tunneling-width matrices $\Gamma^{\alpha}$.
We calculate the transport properties using nonequilibrium Green's functions. 
The retarded Green's function of the junction is given by $G^{r}(\omega)=[\mathbb{S}\omega-H_{\rm gra}-\Sigma_{T}(\omega)]^{-1}$, where 
$\Sigma_{T}=-i\sum_{\alpha}\Gamma^{\alpha}/2$ is the tunneling self-energy.
We take the contact-system couplings in the broad-band limit, i.e., $\Gamma^{\alpha}_{nm}(\omega)=\Gamma^{\alpha}_{nm}(\mu_{0})$ where
$\mu_{0}$ is the Fermi energy of the metal leads. We also take the contact-system couplings to be diagonal matrices
$\Gamma^{\alpha}_{nm}(\omega)=\sum_{l\in\alpha}\Gamma_{\alpha}\delta_{nl}\delta_{ml}$ coupled to $\pi$-orbitals $n,m$ of the graphene system.
The nonzero elements of $\Gamma^{\alpha}$ ($\alpha=\{\rm {blue,\ red}\}$) are at sites
indicated by either a blue or red square in figs.\ \ref{fig:TemperatureBias} and \ref{fig:VoltageBias}, 
corresponding to the carbon atoms of graphene covalently bonded to the contact reservoirs. The tunneling matrix element at each coupling site is set as $\Gamma_{\alpha}=0.5\ \rm {eV}$
for both the contacts (blue and red).
$\mathbb{S}$ is the overlap-matrix between the atomic orbitals on different sites and we take $\mathbb{S}=\mathbb{I}$, 
i.e., an orthonormal set of atomic orbitals.
The tunneling-width matrix $\Gamma^{p}$ describing the probe-sample coupling is also treated in the broad-band limit.
The probe is in the tunneling regime
and the probe-system coupling is weak (few meV) in comparison to the system-reservoir couplings. 

\subsection*{Probe-Sample Coupling}

The scanning tunneling thermometer is modeled as an atomically sharp Pt tip operating in the tunneling regime at a height
of $3\ \rm{\AA}$ above the plane of the carbon nuclei in graphene.
The probe tunneling-width matrices may be described in general
as \cite{Bergfield2012}
$\Gamma^{p}_{nm}(\omega)=2\pi\sum_{l\in \{s,p,d ...\}} C_{l} V^{n}_{l}V^{n^*}_{l}\rho_{l}^{p}(\omega)$, where $\rho_{l}^{p}(\omega)$ 
is the local density of states of the apex atom
in the probe electrode and $V^{m}_{l}$,$V^{n}_{l}$ are the tunneling matrix elements between the l-orbital of the apex atom in
the probe and the $m^{th}$, $n^{th}$ $\pi$-orbitals in graphene. The constants $C_{l}=C\ \forall l$ and has been determined
by matching with the peak of the experimental conductance histogram \cite{Kiguchi08}. 
We consider the Pt tip to be dominated by the d-orbital character (80\%) although
other contributions (s -- 10\% and p -- 10\%) are also taken as described
in Ref.\ \cite{Bergfield2012}. In the calculation of the tunneling matrix elements, 
the $\pi$-orbitals of graphene are taken to be hydrogenic $2p_{z}$ orbitals with an effective nuclear charge $Z=3.22$ \cite{Barr2012}.
The tunneling-width matrix $\Gamma^{p}$ describing
the probe-system coupling is
in general non-diagonal.

\section*{Acknowledgements}

The authors gratefully acknowledge useful discussions with Brian J. LeRoy and Oliver L.A. Monti. This work was supported by the U.S.
Department of Energy (DOE), Office of Science, under Award No. DE-SC0006699.
% \section*{Meeting with Brian LeRoy, Aug 23rd, 2018}
% 
% Equilibration times --- Be specific. 
% Make the point that the specific system considered is not going to affect the generality of our results: Our results are completely
% general!
% Crossno2016: Breakdown of WF Law. Comment on this paper. They specifically chose a regime where the WF law breaks
% since the thermal energy is comparable to the fermi energy.

\bibliographystyle{unsrt}
\bibliography{./refs}
\end{document}

\title{Supplementary Information}
\author[1,2]{Abhay Shastry\thanks{Corresponding Author: abhayshastry@email.arizona.edu}}
\author[1,3]{Sosuke Inui}
\author[1]{Charles\ A.\ Stafford}
\affil[1]{Department of Physics, University of Arizona, Tucson, AZ 85721, USA}
\affil[2]{Chemical Physics Theory Group, Department of Chemistry, University of Toronto, 80 St. George Street, Toronto, ON M5S 3H4, Canada}
\affil[3]{Department of Physics, Osaka City University, Sugimoto 3-3-138, Sumiyoshi-Ku, Osaka 558-8585, Japan}
\maketitle
\section{Elastic Transport}

We explicitly show the derivation of the Wiedemann-Franz law for elastic transport below. 
The steady-state currents flowing into reservoir $p$, through a quantum conductor where elastic processes dominate the transport,
can be written in a form analogous to the multiterminal B{\"u}ttiker formula \cite{Shastry2015}
\begin{equation}
I_{p}^{(\nu)}= \frac{1}{h} \sum_{\alpha}\int_{-\infty}^\infty d\omega\; (\omega-\mu_p)^\nu \, 
%{\rm Tr}\left\{
{\myT}_{p\alpha}(\omega) %\right\}
\left[f_\alpha(\omega)-f_p(\omega)\right],
\label{ElasticTransport}
\end{equation}
where
\begin{equation}
{\myT}_{p\alpha}(\omega)={\rm Tr}\left\{\Gamma^p(\omega) G^r(\omega) \Gamma^\alpha(\omega) G^a(\omega)\right\}
\label{TMatrix}
%\label{eq:transmission}
\end{equation}
is the transmission function for an electron originating in reservoir $\alpha$ to tunnel into reservoir $p$. 
Our notation
uses $\nu=0$ to refer to the particle current and $\nu=1$ to refer to the electronic contribution to the heat
current. $G^{r}\ (G^{a})$ is the retarded (advanced) Green's function. $\Gamma^{p}$ and $\Gamma^{\alpha}$
are the tunneling width matrices describing the coupling of the system to the probe and contact $\alpha$ respectively.
The main article gives expressions in terms of the {\em electrical current} $I_{p}$ which
is related to the particle current by
\begin{equation}
\begin{aligned}
I_{p}=& - e I^{(0)}_{p}
     =& - \frac{e}{h} \sum_{\alpha}\int_{-\infty}^\infty d\omega\;  \, 
{\myT}_{p\alpha}(\omega) %\right\}
\left[f_\alpha(\omega)-f_p(\omega)\right],
\end{aligned}
\end{equation}
whereas the electronic heat current is simply 
\begin{equation}
J_{p}= I^{(1)}_{p}.
\end{equation}

Operation within the linear response regime allows one to expand the fermi functions $f_{\alpha}$ and 
$f_{p}$ to linear order near the equilibrium temperature and chemical potential
\begin{equation}
\begin{aligned}
f_{\alpha}-f_{p}=&  \bigg(\frac{\partial f}{\partial \mu}\bigg)\bigg|_{\mu_{0},T_{0}}(\mu_{\alpha}- \mu_{p})
                  + \bigg(\frac{\partial f}{\partial T}\bigg)\bigg|_{\mu_{0},T_{0}}(T_{\alpha}-T_{p})\\
=&  \bigg(-\frac{\partial f}{\partial\omega}\bigg)\bigg|_{\mu_{0},T_{0}}\, \big(-e(V_{\alpha}-V_{p})\big)
 + (\omega - \mu_{0})\bigg(-\frac{\partial f}{\partial\omega}\bigg)\bigg|_{\mu_{0},T_{0}}\, \frac{(T_{\alpha}-T_{p})}{T_{0}}.
\end{aligned}
\label{fermiExpansion}
\end{equation}
The electrical current
\begin{equation}
I_{p} =\sum_{\alpha}\ \Lfun{0}{p\alpha} (V_{\alpha}-V_{p}) + \Lfun{1}{p\alpha} \frac{(T_{\alpha}-T_{p})}{T_{0}},
\end{equation}
to linear order in the voltage and temperature gradients.
Using Eq.\ (\ref{fermiExpansion}) in Eq.\ (\ref{ElasticTransport}), we obtain the expressions for the linear response
coefficients 
\begin{equation}
\Lfun{0}{p\alpha} = \frac{e^{2}}{h} \int_{-\infty}^\infty d\omega\; \myT_{p\alpha}(\omega)
\bigg(-\frac{\partial f}{\partial\omega}\bigg)\bigg|_{\mu_{0},T_{0}}
\end{equation}
and
\begin{equation}
\Lfun{1}{p\alpha} = \frac{-e}{h} \int_{-\infty}^\infty d\omega\; (\omega- \mu_{0})\,\myT_{p\alpha}(\omega)
\bigg(-\frac{\partial f}{\partial\omega}\bigg)\bigg|_{\mu_{0},T_{0}}.
\end{equation}
The heat current 
\begin{equation}
J_{p}= \sum_{\alpha}\ \Lfun{1}{p\alpha} (V_{\alpha}-V_{p}) + \Lfun{2}{p\alpha} \frac{(T_{\alpha}-T_{p})}{T_{0}},
\end{equation}
where we have taken $\mu_{p}\approx \mu_{0}$ in Eq.\ (\ref{ElasticTransport}) since we are interested in terms up to
the linear order.
Again, we infer from Eqs.\ (\ref{fermiExpansion}) and (\ref{ElasticTransport}) that
\begin{equation}
\Lfun{2}{p\alpha} = \frac{1}{h} \int_{-\infty}^\infty d\omega\; (\omega- \mu_{0})^{2}\,\myT_{p\alpha}(\omega)
\bigg(-\frac{\partial f}{\partial\omega}\bigg)\bigg|_{\mu_{0},T_{0}}.
\end{equation}

The derivative of the fermi function appears in the expressions for all the linear response coefficients and
we may use the Sommerfeld series expansion \cite{Shastry2015,AshcroftAndMerminBook}. We find that
\begin{equation}
\begin{aligned}
\frac{h}{e^{2}}\Lfun{0}{p\alpha} = \myT_{p\alpha}(\mu_{0})\ +& 2\ \Theta(2)(k_{B}T_{0})^{2}\myT_{p\alpha}^{(2)}(\mu_{0})\ \ \ \ \ \\
+& 2\ \Theta(4) (k_{B}T_{0})^{4}\myT_{p\alpha}^{(4)}(\mu_{0}) + ...
\end{aligned}
\label{L0Series}
\end{equation}
and
\begin{equation}
%\nonumber
\begin{aligned}
-\frac{h}{e}\Lfun{1}{p\alpha}= 4\ \Theta(2)(k_{B}T_{0})^{2}\myT_{p\alpha}^{(1)}(\mu_{0})&+ 
8\ \Theta(4)(k_{B}T_{0})^{4}\myT_{p\alpha}^{(3)}(\mu_{0})\\
&+12\ \Theta(6)(k_{B}T_{0})^{6}\myT_{p\alpha}^{(5)}(\mu_{0})+...
\end{aligned}
\label{L1Series}
\end{equation}
and
\begin{equation}
%\nonumber
\begin{aligned}
h\Lfun{2}{p\alpha}=4\ \Theta(2)(k_{B}T_{0})^{2}\myT_{p\alpha}(\mu_{0})&+ 24\ \Theta(4)(k_{B}T_{0})^{4}\myT_{p\alpha}^{(2)}(\mu_{0})\\
&+60\ \Theta(6)(k_{B}T_{0})^{6}\myT_{p\alpha}^{(4)}(\mu_{0})+...,
\end{aligned}
\label{L2Series}
\end{equation} 
where
we use the notation from ref.\ \cite{Shastry2015}:
$\myT^{(k)}_{p\alpha}(\mu_{0})$ denotes the k-th derivative of the transmission function 
$\myT_{p\alpha}(\omega)$ at $\omega=\mu_{0}$ and
$\Theta$ is a numerical factor related to the Riemann-Zeta function
         \begin{equation}
         \Theta(k+1)= \big(1- \frac{1}{2^{k}}\big)\zeta(k+1).
         \end{equation}
Explicitly:
         \begin{equation}
         \begin{aligned}
         \Theta(2)&= \frac{\pi^{2}}{12}\\
         \Theta(4)&= \bigg(\frac{7}{8}\bigg)\frac{\pi^{4}}{90}\\
         \Theta(6)&= \bigg(\frac{31}{32}\bigg)\frac{\pi^{6}}{945}.
         \end{aligned}
         \end{equation}

The transmission function has appreciable changes on an energy scale determined by the system's Hamiltonian and its
couplings to the contacts. We thus define the characteristic energy scale $\Delta$
\begin{equation}
\myT_{p\alpha}(\mu_{0})= \Delta^{2} \myT_{p\alpha}^{(2)}(\mu_{0}),
\end{equation}
which is typically much larger than the thermal energy $k_{B}T_{0}$ for most experimental setups.

The following relation connecting $\Lfun{0}{p\alpha}$ and $\Lfun{2}{p\alpha}$,
from Eqs.\ (\ref{L0Series}) and (\ref{L2Series}), is the Wiedemann-Franz law:
\begin{equation}
\Lfun{2}{p\alpha}=\frac{\pi^{2}k_{B}^{2} T_{0}^{2}}{3e^{2}}\Lfun{0}{p\alpha}\bigg(1 + \frac{8\pi^{2}}{15}\bigg(\frac{k_{B}T_{0}}{\Delta}\bigg)^{2}
 + ... \bigg).
\end{equation}

% \begin{equation}
% \Lfun{2}{p\alpha}=\frac{\pi^{2}k_{B}^{2} T_{0}^{2}}{3e^{2}}\Lfun{0}{p\alpha}\bigg(1 +\mathcal{O}\big((k_{B}T_{0}/\Delta)^{2}\big)\bigg).
% \end{equation}
\section{Calibration of Temperature}

\begin{figure}[tbh]
\centering
\captionsetup{justification=raggedright,
singlelinecheck=false
}

{\includegraphics[width=2.5in]{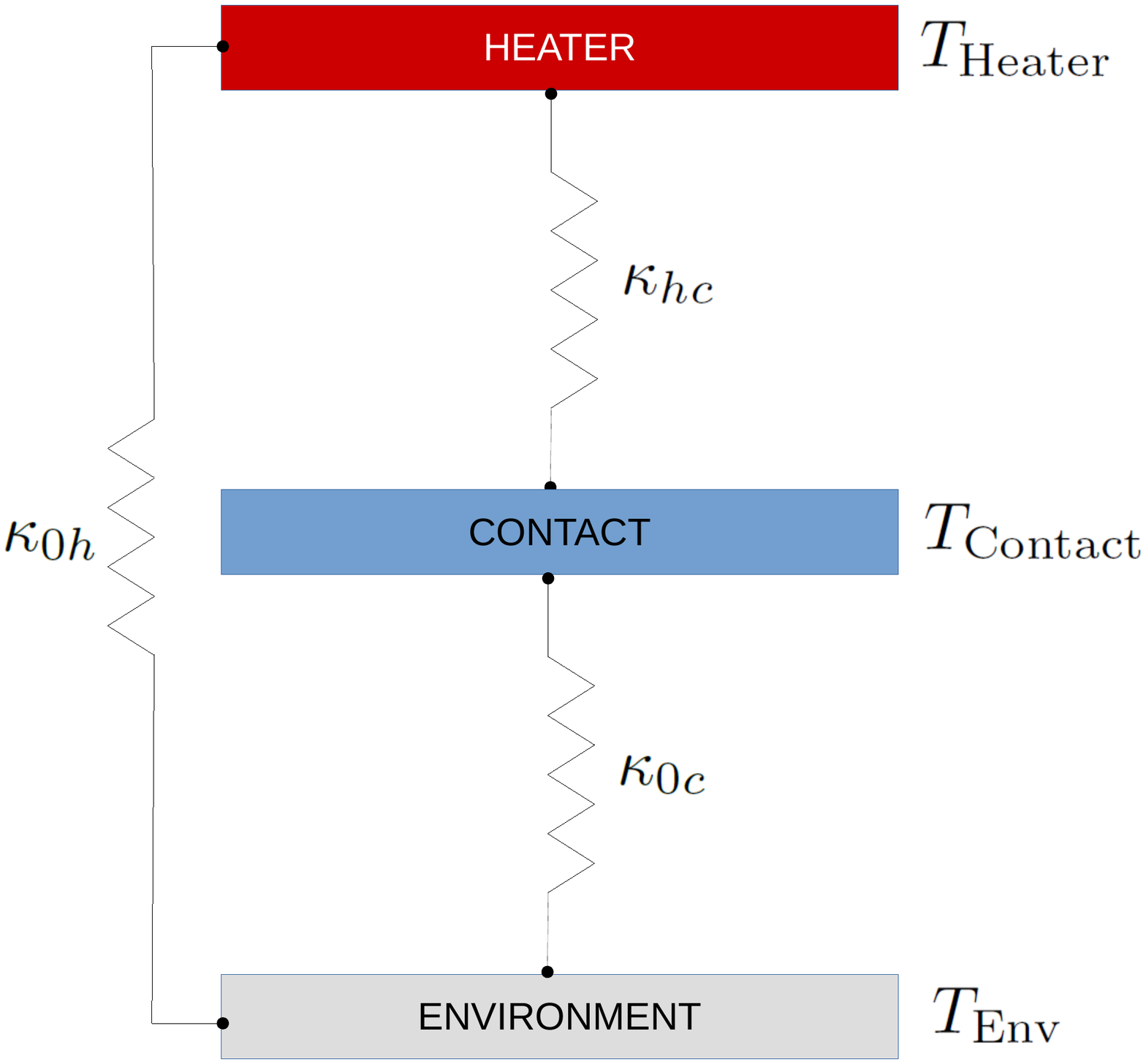}}
\caption{Thermal circuit for heat transfer between the Pt-heater, the metal contact, and the ambient environment. 
The contact temperature is nearly equal to that of the Pt-heater, $T_{\rm{Contact}}\approx T_{\rm{Heater}}$, 
when they are in good thermal contact $\kappa_{hc}\gg\kappa_{0c}$.
}
\label{fig:ThermalCircuit}
\end{figure}

The thermoelectric circuit requires the calibration of the contact temperatures which we describe here. 
The Pt-heater is fabricated atop an electrically insulating layer above the metal contact $\alpha$ and has a thermal 
conductivity $\kappa_{hc}$ with the contact. The temperature of the Pt-heater is inferred from its resistivity. The contact $\alpha$
is heated when an electrical current is passed in the Pt-heater but it also loses heat to the ambient environment
which is at the equilibrium temperature $T_{\rm{Env}}=T_{0}$. We denote the
thermal conductivity between the contact and the ambient environment by $\kappa_{0}$. The thermal circuit is shown 
in Fig.\ \ref{fig:ThermalCircuit}.

The heat current flowing into the contact is given by
\begin{equation}
\dot{Q}_{\rm{in}} = \kappa_{hc} \big(T_{\rm{Heater}}-T_{\rm{Contact}}\big)
\end{equation}
whereas the heat current flowing out 
\begin{equation}
\dot{Q}_{\rm{out}} = \kappa_{0c} \big(T_{\rm{Contact}}-T_{\rm{Env}}\big).
\end{equation}
In steady state, the rate of heat flow into the contact is equal to the rate of heat lost to the ambient environment
and we find
\begin{equation}
T_{\rm{Contact}}= \frac{\kappa_{hc}T_{\rm{Heater}} + \kappa_{0c} T_{\rm{Env}}}{\kappa_{hc} + \kappa_{0c}}.
\end{equation}
When the heater is in good thermal contact $\kappa_{hc}\gg\kappa_{0c}$, we find that
\begin{equation}
T_{\rm{Contact}}\approx T_{\rm{Heater}}.
\end{equation}

An alternating voltage 
$V(t)= V_{\rm{max}}\cos(\omega t)$
results in a current $I(t)= G_{\rm{Pt}} V(t)$ in the heater. The power dissipated via Joule heating is given by
\begin{equation}
P = G_{\rm{Pt}} V_{\rm{max}}^{2}\cos^{2}(\omega t)
  = \frac{1}{2} G_{\rm{Pt}} V_{\rm{max}}^{2}\big(1 + \cos(2\omega t) \big),
\end{equation}
which results in $2\omega$ modulations of the heater temperature
\begin{equation}
T_{\rm{Heater}} = T_{0} + \Delta T_{\rm{max}} \big(1 + \cos(2\omega t)\big),
\label{eq:HeaterTemperatureModulation}
\end{equation}
since the net power dissipated by the heater can be written as
\begin{equation}
\begin{aligned}
P =&\ \kappa (T_{\rm{Heater}}- T_{0}),\ \text{where},\\
\kappa =&\ \kappa_{0h} + \frac{\kappa_{hc} \ \kappa_{0c}}{\kappa_{hc} + \kappa_{0c}}, 
\end{aligned}
\end{equation}
as seen from the thermal circuit shown in fig.\ \ref{fig:ThermalCircuit}.

The temperature of the heater is inferred from the conductance (or resistivity) dependence of the Pt heating element $G_{\rm{Pt}}(T)$.
The modulation frequency must be chosen so that $\omega\ll 1/\tau$, where $\tau$ is the thermal time constant of the metal contact,
so that it has enough time to thermally equilibrate. It is of course understood that such a frequency allows the heater itself to
equilibrate and enter a steady-state of heat transfer with the metal contact. 
The temperature modulations in the metal contact closely follow that of the heater when there is good thermal contact:
\begin{equation}
T_{\rm{Contact}} (t) = T_{0} + \Delta T_{\rm{max}} \big(1 + \cos(2\omega t)\big).
\label{eq:ContactTemperatureModulation}
\end{equation}
We have chosen $\Delta T_{\rm{max}}$ such that the contact would reach a
maximum temperature of $T_{0} + 2\Delta T_{\rm{max}}$. The calibration fixes $\Delta T_{\rm{max}}$ accurately.

We also note that the temperature modulations can be obtained by means other than using a Pt resistor. A graphene flake
itself undergoes Joule heating and could therefore be used as a heating element so long as one is able to calibrate its temperature
accurately.

\section{Tunneling Currents}

\begin{figure*}[tbh]
\centering
\captionsetup{justification=raggedright,
singlelinecheck=false
}
%\hspace{-0.5in}
{\includegraphics[width=2.7in]{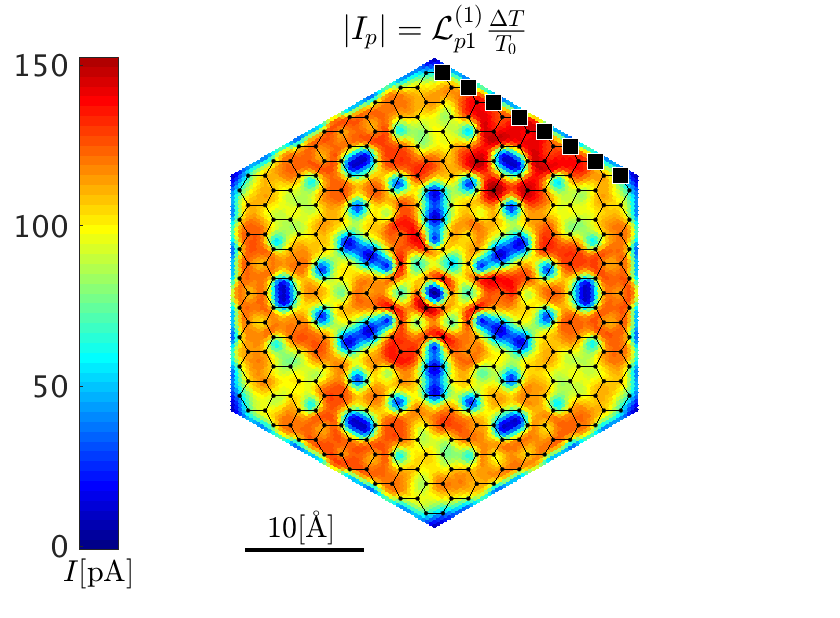}}\hspace{-0.7in}
{\includegraphics[width=2.7in]{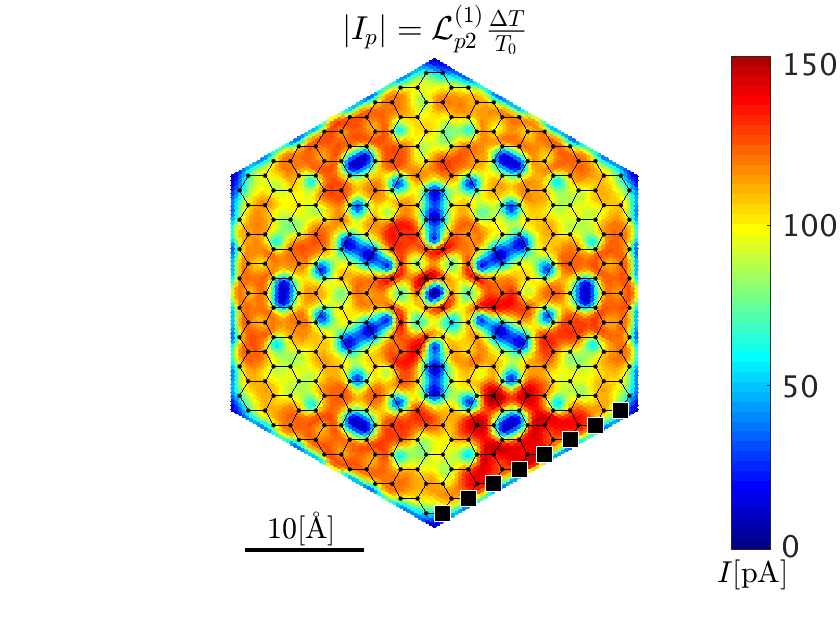}}
\caption{Amplitude of the tunneling current in the thermoelectric circuit. The left (right) panel
shows the tunneling current amplitude resulting from the heating of the first (second) contact
as shown with the black squares in the corresponding panel. The gating potential is set at
$\mu_{0}=-2.28\ \rm{eV}$ with respect to the Dirac point.
The amplitude of temperature variations in the contacts [cf.\ Eq.\ (\ref{eq:ContactTemperatureModulation})]
is taken to be 10\% of the equilibrium temperature $\frac{\Delta T_{\rm{max}}}{T_{0}}= 0.1$. 
}
\label{fig:TunnelingCurrentThermoelectric}
\end{figure*}

\subsection{Thermoelectric Circuit}

The tunneling current resulting from the heating of the contact is given by
\begin{equation}
I_{p}= \Lfun{1}{p\alpha}\frac{(T_{\alpha} -T_{0})}{T_{0}}
\end{equation}
during the operation of the thermoelectric circuit. Standard lock-in techniques are
employed to measure the current amplitude at frequency $2\omega$. It is easy to see from
Eq.\ (\ref{eq:ContactTemperatureModulation}) that the current amplitude 
\begin{equation}
I_{p}\big|_{2\omega} = \Lfun{1}{p\alpha} \frac{\Delta T_{\rm{max}}}{T_{0}}.
\label{eq:CurrentAmplitudeThermoelectric}
\end{equation}
We show the spatial variation of the tunneling current amplitude in fig.\ \ref{fig:TunnelingCurrentThermoelectric}. The probe
is held at a constant height of $3\ \rm{\AA}$ above the plane of the sample. We assume a modest increase in the contact temperature
by setting $\Delta T_{\rm{max}} = (10\%)\ T_{0}$ where the equilibrium temperature $T_{0}= 4\ \rm{K}$. 
The corresponding contact $\alpha=\{1, 2\}$ is shown by black squares in fig.\ \ref{fig:TunnelingCurrentThermoelectric} and represent
the sites of the sample which are covalently bonded to the metal contact $\alpha$. $\alpha=1$ is shown on the left panel and
$\alpha=2$ is shown on the right panel in fig.\ \ref{fig:TunnelingCurrentThermoelectric}.
The tunneling current
amplitude is as high as $150\ \rm{pA}$ at some points on the sample and is therefore well within the reach of present experimental
resolution. Since we illustrate our numerical results for an experiment performed at liquid He temperatures ($4\ \rm{K}$),
the thermoelectric response is suppressed and gating becomes important. If, for example, $T_{0}$ was set to $40\ \rm{K}$,
we would have a hundred-fold increase in the tunneling current amplitude [cf.\ Eqs.\ (\ref{eq:CurrentAmplitudeThermoelectric})
and (\ref{L1Series})] and gating would be less important.

%Above value should be sufficiently large for an interesting experiment blah blah
\subsubsection{Gating}

\begin{figure*}[tbh]
\centering
\captionsetup{justification=raggedright,
singlelinecheck=false
}
\hspace{-0.5in}
{\includegraphics[width=4.5in]{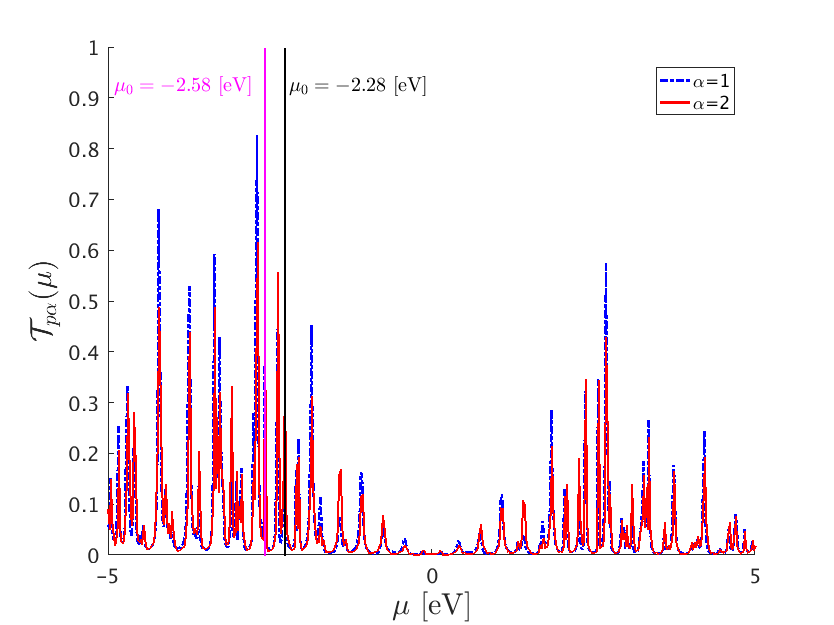}}
\caption{Transmission function $\myT_{p\alpha}$ from contact $\alpha$ into the STM probe $p$. $\alpha =\{1,2\}$ are
shown in (dotted-dashed) blue and red respectively. The conductance circuit measurement is illustrated at a gating potential of
$\mu_{0}= -2.58\ \rm{eV}$ (magenta vertical line).
However, we illustrate the thermoelectric circuit at a gating potential of $\mu_{0}= -2.28\ \rm{eV}$ (black vertical line) since the 
transmission functions show a large change at that choice of gating, thereby resulting in an enhanced thermoelectric effect. 
}
\label{fig:Transmissions}
\end{figure*}
We find that the system has a sufficiently large thermoelectric response at $4\ \rm{K}$, i.e.\ the current amplitude in
Eq.\ (\ref{eq:CurrentAmplitudeThermoelectric}) is experimentally resolvable, when the system is gated appropriately.
Indeed, our method works perfectly well for systems which do not have a good thermoelectric response. In such a case,
$\Lfun{1}{p\alpha}$ would have a low value and would result in a current amplitude which is too small to measure. 
This merely implies that the thermoelectric contribution to the measured temperature is very small --- that is, a
voltage bias within the linear response regime does not lead to measurable differences in temperatures across the sample.
We have chosen the system's gating so that the thermoelectric response is appreciable and there are measurable temperature
differences across the sample even in the case of a voltage bias. We find this latter case more interesting.

The thermoelectric coefficient depends on the transmission derivative [cf.\ Eq.\ (\ref{L1Series})] near the equilibrium
chemical potential. 
In fig.\ \ref{fig:Transmissions},
we show the transmission functions as a function of the chemical potential. The figure shows the transmission spectra into the probe
from the two contacts $\alpha= \{1,2\}$ for one representative point 
on the sample where the probe is held at a height of $3\ \rm{\AA}$ above
the plane of the sample; The transmission spectra would change from point to point on the sample but will roughly resemble
the one in fig.\ \ref{fig:Transmissions}. The contact $\alpha=1$ is shown by in blue (dotted and dashed) whereas $\alpha=2$ 
is shown in red.
% The reader may refer to fig.\ (\ref{fig:TunnelingCurrentThermoelectric}) where red squares represent
% the sites of the sample covalently bonded to the corresponding contact $\alpha$ --- $\alpha= 1$ is shown on the left panel
% and $\alpha=2$ is shown on the right panel in fig.\ (\ref{fig:TunnelingCurrentThermoelectric}).
We found that the transmission derivatives are enhanced when the chemical 
potential is tuned (via the gate voltage) to $\mu_{0}=-2.28\ \rm{eV}$ and therefore illustrated the thermoelectric circuit
for this choice of gating. The resulting temperature measurement is shown in Fig. 3 of the main paper for a pure voltage bias.
The spatial variations in the transmission derivatives would resemble
the pattern shown in fig.\ \ref{fig:TunnelingCurrentThermoelectric} [cf.\ Eq.\ (\ref{L1Series})].

\subsection{Conductance Circuit}

\begin{figure*}[tbh]
\centering
\captionsetup{justification=raggedright,
singlelinecheck=false
}
%\hspace{-0.5in}
{\includegraphics[width=2.7in]{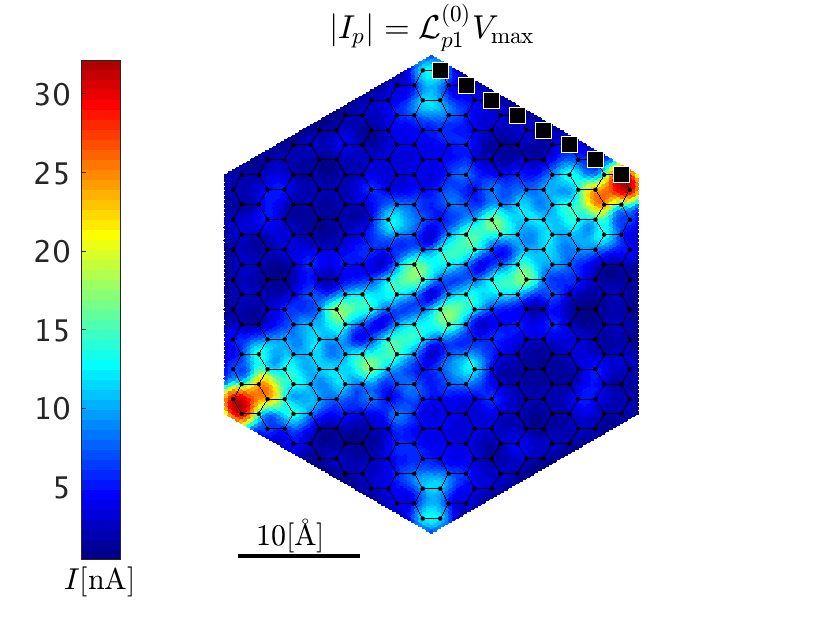}}\hspace{-0.7in}
{\includegraphics[width=2.7in]{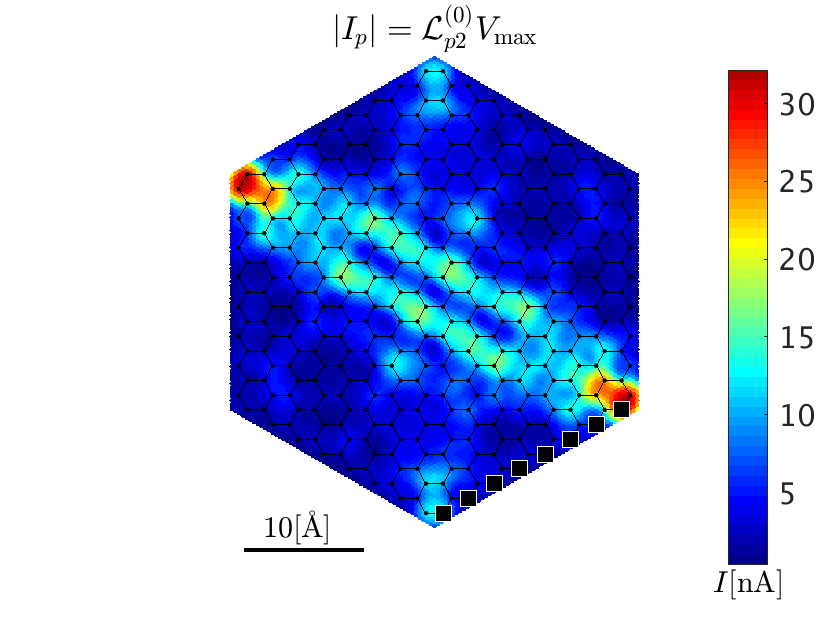}}
\caption{Amplitude of the tunneling current in the conductance circuit. 
The gating potential is set at
$\mu_{0}=-2.58\ \rm{eV}$ with respect to the Dirac point. The left (right) panel
shows the tunneling current amplitude resulting from the voltage bias between the first (second) contact
and the probe, $V(t) = V_{\alpha} - V_{p} = V_{\rm{max}} \cos(\omega t)$,
as shown with the black squares in the corresponding panel. $V_{\rm{max}}=1\ \rm mV$.
}
\label{fig:TunnelingCurrentConductance}
\end{figure*}

The tunneling current resulting from the conductance circuit would simply be
\begin{equation}
I_{p}= \Lfun{0}{p\alpha}(V_{\alpha}-V_{p}).
\end{equation}
We apply an ac voltage $V_{\alpha}-V_{p}= V(t) = V_{\rm{max}}\cos(\omega t)$ across the contact-probe junction
and measure the resulting tunneling current
\begin{equation}
I_{p}(t)=  \Lfun{0}{p\alpha} V_{\rm{max}} \cos(\omega t)
\end{equation} 
using standard lock-in techniques. The tunneling current amplitude at frequency $\omega$
\begin{equation}
I_{p}\big|_{\omega} = \Lfun{0}{p\alpha} V_{\rm{max}}
\end{equation}
is measured across the sample as shown in fig.\ \ref{fig:TunnelingCurrentConductance}. We 
set the amplitude of voltage modulations $V_{\rm{max}}=1\ \rm{mV}$ and a scan of the sample 
is obtained by maintaining the probe tip at a height of $3\ \rm{\AA}$ above the plane of the sample. 
% The gating doesn't play as important a role in the measurement of the conductance since we obtain 
The tunneling current amplitude is as high as $30\ \rm{nA}$ for some regions in the sample. Generally, gating
doesn't play as important a role in the measurement of conductances since we obtain tunneling currents of the order of a few
$\rm{nA}$ for most choices of gating.
The corresponding contact $\alpha=\{1,2\}$ is shown by black squares in fig.\ \ref{fig:TunnelingCurrentConductance} and represent
the sites of the sample which are covalently bonded to the metal contact.  $\alpha=1$ is shown on the left panel and
$\alpha=2$ is shown on the right panel in fig.\ \ref{fig:TunnelingCurrentConductance}. 
The resulting temperature measurement is shown in Fig. 2 of the main paper
for a pure temperature bias. 

\bibliographystyle{unsrt}
\bibliography{./refs}